\documentstyle[preprint,aps,epsf]{revtex}

\newcommand{\newc}{\newcommand}
\newc{\beq}    {\begin{equation}}
\newc{\eeq}    {\end{equation}}
\newc{\beqa}    {\begin{eqnarray}}
\newc{\eeqa}    {\end{eqnarray}}

\def\PLA{{\em Phys. Lett.}  {\bf A}}
\def\PRL{{\em Phys. Rev. Lett. }}

\def\PRD{{\em Phys. Rev.} {\bf D} }
\def\PR{{\em Phys. Rev.}  }

\begin{document}


%
\title{ Stochastic Processes and the Dirac Equation
 with External Fields }
\author{ Jae-weon Lee  and Eok Kyun Lee  }
\address{
 Department of Chemistry,  School of Molecular Science (BK 21),
Korea Advanced   Institute of Science and Technology,  Taejon
 305-701, Korea.}

\author{ Hae Myoung Kwon and In-gyu Koh
 }
\address{
Department of Physics, Korea Advanced Institute of Science and
Technology, Taejon 305-701, Korea}

\author{ Yeong Deok Han }
\address{Department of Physics,  Woosuk University, 490
 Hujeong-ri, Samrye-eup, Wanju-kun, Cheonbuk, Korea }

\maketitle

\vspace{1cm}

 The equation describing the stochastic motion of a
classical particle in 1+1-dimensional space-time is connected to
the Dirac equation with external gauge fields. The effects of assigning
different turning probabilities to the forward and the backward
moving particles in time are  discussed.

\pacs{PACS: 03.65.Ca, 02.30.+g, 05.40.+j  }


For many decades, the relations between stochastic processes and
quantum mechanics have attracted much
attention\cite{nelson,jacobson,jacobson2}.
The Schr$\ddot{o}$dinger's equation for a nonrelativistic
electron
and the heat equation are related\cite{wiener,schulman} as the
 Feynman path integral\cite{feynman} and the Wiener integral
are connected by an analytic continuation.
This correspondence was extended to the relativistic
case by studying the telegrapher's equation
which produced  the Dirac equation when analytically continued
\cite{gaveau}.
A decade ago, McKeon and Ord further extended the study
by introducing the probability of moving backward in time and
deriving the Dirac equation in 1+1-dimensional space-time
without analytical continuation \cite{mckeon}.

 In this paper, we will extend these considerations by introducing
external gauge  fields and giving different `turning'
probabilities to forward and backward moving particles in time.
In these models, particles are supposed to
suffer random, Poisson-distributed
reversals in the moving direction.
Notations and
 procedures similar to those used in Ref. 8 are adopted.
Let $F_+~(F_-)$ and $B_+~(B_-)$ be the probabilities for
 moving to the positive(negative)
$x$-direction
for  forward-moving particles  and backward-moving particles,
respectively,
in the $t$-direction.
The probability for  reversal in time interval $\triangle t$ is
$a_{R}^F~(a_{L}^F) \triangle t$ for turning right(left)
with respect to the  direction of motion  for
the forward moving particles, and $a^B_{R}(a^B_{L})\triangle t$
 for the backward moving particles(See Fig.~1).
In this paper, we assume that $a^{F,B}_{R,L}$ can be
analytically continued to the imaginary part.
This requires a non-trivial interpretation against the positiveness
of the  probability.

When there is an external field,
it is reasonable to assume that the forward and the backward moving
particles have different turning probabilities, i.e.,
$a^F_{R,L} \neq a^B_{R,L}$.
From Fig.~1, it is easy to derive
the master equation describing the evolution
of the probability $F_\pm$ on the $x-t$ plane:
\beqa
F_\pm(x,t)&=&(1-a^F_L\triangle t -a^F_R \triangle t) \nonumber\\
&F_\pm&(x\mp \triangle x, t-\triangle t) \nonumber \\
&+&a^B_{L,R}B_\pm \triangle t(x\mp \triangle x, t+\triangle t)\nonumber\\
&+&a^F_{R,L}F_\mp \triangle t(x\pm \triangle x, t-\triangle t).
\label{F1}
 \eeqa
Then, the `causality condition' used in
Ref. 8 that $F_\pm(x,t)=B_\mp(x\pm \triangle x,
t+\triangle t)$ gives  similar equation for $B_\pm$:
\beqa
B_\pm(x&\mp& \triangle x, t+\triangle t)
= F_\mp(x,t) \nonumber \\
&=&(1-a^F_L\triangle t -a^F_R \triangle t)
 B_\pm(x,t) \nonumber\\
 &+&a^B_{R,L}\triangle t F_\pm(x, t)
+a^F_{L,R}\triangle t B_\mp(x, t).
\label{B}
 \eeqa

Expanding Eq.(\ref{F1}) and Eq.(\ref{B})
 to first order in $\triangle x$ and $\triangle t$
leads to two differential equations:
\beqa
\pm v \frac{\partial
F_\pm}{\partial x}+ \frac{\partial F_\pm}{\partial t}&=&
-a^F_{L,R}F_\pm +a^B_{L,R}B_\pm \nonumber\\
&-& a^F_{R,L}F_\pm +a^F_{R,L}F_\pm
\label{Fpm}
\eeqa

 and

\beqa
\pm  v\frac{\partial B_\mp}{\partial x}+
\frac{\partial B_\mp}{\partial t}&=& -a^F_{L,R}B_\mp +a^B_{L,R}F_\mp
\nonumber \\
&-& a^F_{R,L}B_\mp +a^F_{R,L}B_\pm,
 \label{Bpm}
 \eeqa
  where $v\equiv
dx/dt$.
Subtracting Eq.~(\ref{Bpm}) from Eq.~(\ref{Fpm}) yields
\beqa
 v\frac{\partial (F_\pm - B_\mp)}{\partial x} &\pm&
\frac{\partial( F_\pm - B_\mp)}{\partial t}= \nonumber \\
&\mp&(a^F_{L,R}+a^F_{R,L})(F_\pm-B_\mp) \nonumber \\
&\pm& (a^F_{R,L}-a^B_{L,R})(F_\mp-B_\pm).
\label{FB}
\eeqa
Now, we will show that this equation can
be related to the 1+1-dimensional Dirac equation
with external fields.

First, let us consider the case where
  the external fields are gauge fields.
The Dirac equation for electrons with $U(1)$ gauge fields is
\beq
\left ( i\gamma^\mu(\hbar \partial_\mu -ieA_\mu)-mc^2 \right
)\psi=0.
\label{dirac}
 \eeq
 As is well known, by a gauge transformation
$A_1\rightarrow A_1+\partial_1 \Lambda$ with a function $\Lambda$
satisfying $\partial_\mu\partial^\mu \Lambda=0$ in the Lorentz gauge
$\partial^\mu A_\mu=0$, $A_1$ can be always chosen as zero.
Then, in $1+1$ dimensions, the Dirac equation can be written  as
\beq
i\hbar\partial_t\psi+ic\hbar\sigma_z \partial_x \psi = -eA_0\psi+
 mc^2 \sigma_y \psi,
\label{dirac2}
\eeq
where $\gamma^0=\sigma_y,\gamma^1=i\sigma_x$, and $\sigma_x$
and $\sigma_y$
are the Pauli matrices.
Substituting a two-dimensional spinor $\psi^T\equiv(\psi_+,\psi_-)$ into
Eq.~(\ref{dirac2}) gives
\beq
c\partial_x\psi_\pm \pm \partial_t \psi_\pm = \pm
\frac{ieA_0}{\hbar}\psi_\pm
- \frac{ mc^2}{\hbar} \psi_\mp.
\label{dirac3}
\eeq
Then, using  the identities
\beqa
v&=&c,\nonumber\\
  F_\pm-B_\mp &=& \psi_\pm ,\nonumber\\
(a^F_{R}-a^B_{L}) &=& -(a^F_{L}-a^B_{R})
=-mc^2/\hbar,
\label{F}
\eeqa
\beq
(a^F_{L}+a^F_{R})=-ieA_0/\hbar,
\label{A}
\eeq
one easily finds that Eq.~(\ref{FB}) can be reduced
to Eq.~(\ref{dirac3}),
 the 1+1-dimensional Dirac equation
 with external gauge fields.
Note that from Eq.~(\ref{F}),
 $a^F_{R}+a^F_{L}$ should be equal to $a^B_{R}+a^B_{L}$
to guarantee a unique mass $m$ for the particle.
Obviously, this relation is automatically satisfied when
$a^{F}_{R,L}=a^{B}_{R,L}\equiv a_{R,L}$ as in Ref. 8,
where $\psi_\pm$ was defined as $exp\{(a_L+a_R)t\} (F_\pm-B_\mp)$
rather than $(F_\pm-B_\mp)$ itself. Since the gauge fields contribute to
the phase of the matter fields as $exp\{ie\int dx^\mu A_\mu/\hbar \}$,
it seems to be reasonable to take $a_L+a_R$ to be proportional to
$A_0$. Furthermore, when $a^{F}_{R,L}=a^{B}_{R,L}\equiv a_{R,L}$,
 one can obtain $a_R$ and $a_L$ explicitly
from Eq.~(\ref{F}) and Eq.~(\ref{A}), $i.e.$,

\beqa
a_R&=&\frac{-ieA_0-mc^2}{2\hbar}, \nonumber \\
a_L&=&\frac{-ieA_0+mc^2}{2\hbar}.
\label{arl}
\eeqa

If the gauge fields are space-time dependent,
 the $a^{F,B}_{R,L}$ are not constants, but functions of $x,t$.
However,  including this space-time dependency of $a^{F,B}_{R,L}$ in
Eq.~(\ref{F}) and Eq.~(\ref{B}), for example,  by replacing $a^F_R$ by
$a^F_R(x+\triangle x,
t-\triangle t)$, does not change the
results because it gives the second-order
terms when expanded in $\triangle x$ and $\triangle t$.
As one can see in Eq.(\ref{arl}),
 the $a^{F,B}_{R,L}$ are complex numbers when
 the external  fields are gauge fields.
This means that we need an analytic continuation, which was
already noted in Ref. 8.
This is unavoidable  because the terms
$\hbar \partial_\mu -ieA_\mu$ in the Dirac equation
 always give a
 factor of $i$ for $A_\mu$ in comparison with $\partial_\mu$.

In summary, the relation between
the stochastic motion and the Dirac equation with external
gauge fields
in 1+1-dimensional space-time is studied.
We also investigated  the effects of assigning  different
turning
probabilities to the
particles moving forward and  backward  in time.

\vskip 1cm
 ACKNOWLEDGMENT
\vskip .3 cm
 This work was supported by BK21.

\newpage

 FIGURES \\
 Fig. 1. Schematic diagram showing the stochastic motion of
particles in the $x-t$ plane and the definitions
of the probabilities.
The center dot denotes the point $(x,t)$.

\newpage


\begin{figure}
\epsfysize=8cm \epsfbox{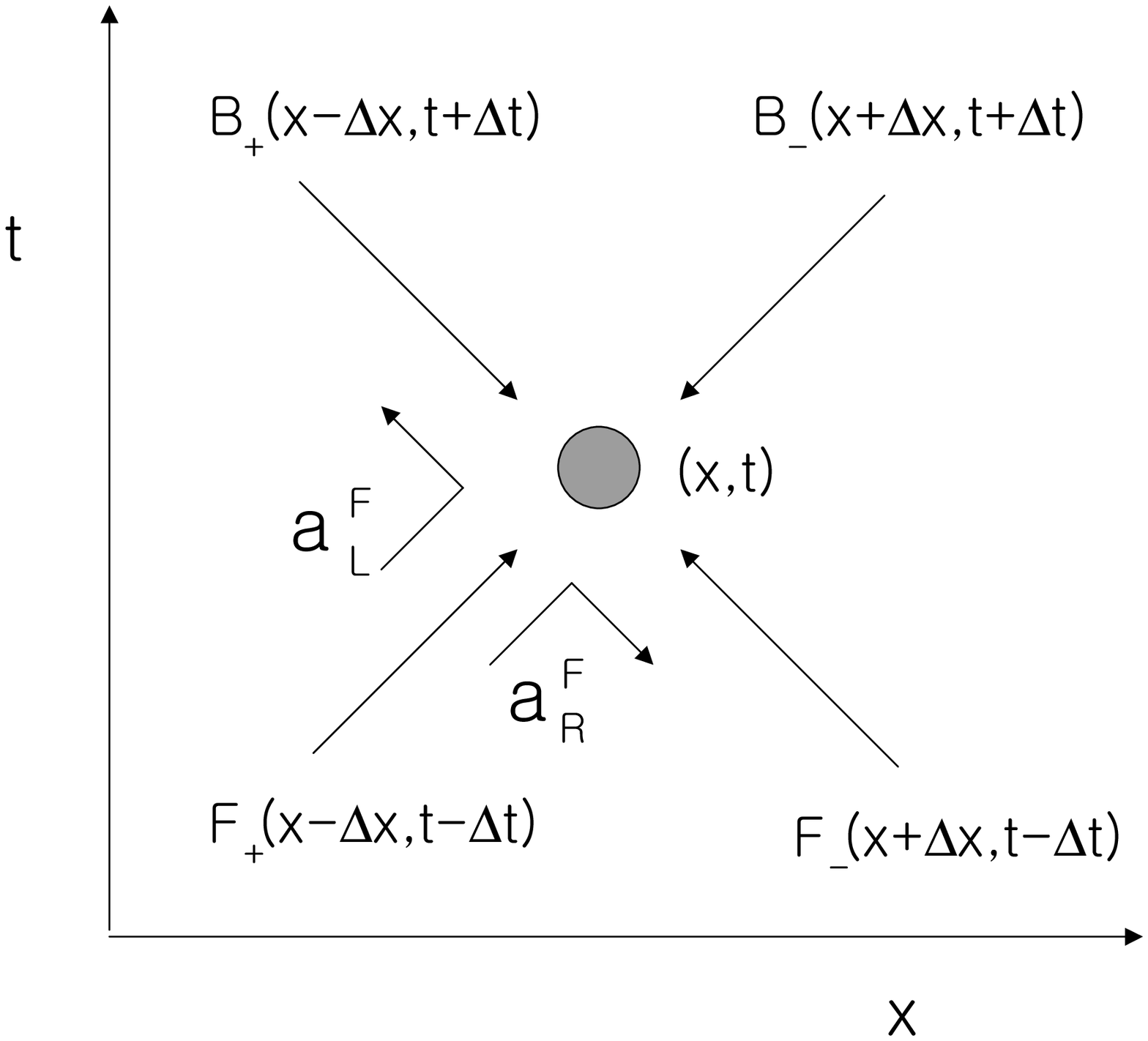}
\vspace{5cm}
\caption{\label{fig1}}

\end{figure}

\end{document}